\documentclass[fleqn,10pt]{wlscirep}

\usepackage[utf8]{inputenc}
\usepackage[T1]{fontenc}
\usepackage{graphicx}  
\usepackage{dcolumn}   
\usepackage{bm}        
\usepackage{amssymb}  
\usepackage{amsmath}
\usepackage{hyperref,color,braket}
\usepackage{textcomp} 
\usepackage{calrsfs} 
\usepackage{mathrsfs} 
\usepackage{xcolor}

\DeclareUnicodeCharacter{03D5}{\phi}

\begin{document}
\title{\texorpdfstring{Modeling Realistic Dynamics of Nanoparticle Dimers for Magneto-Optical Matter}}

\author[1-3,*]{Ricardo Mart\'{\i}n Abraham-Ekeroth}
\author[1]{Dani Torrent}
\affil[1]{GROC-UJI, Institute of New Imaging Technologies, Universitat Jaume I, Av. De Vicent Sos Baynat s/n, 12071 Castell\'{o}n, Spain}
\affil[2]{IFAS, Institute of Physics "Arroyo Seco", National University of Central Buenos Aires, Pinto 399, 7000 Tandil, Argentina}
\affil[3]{Center of Research of Physics and Engineering of Central Buenos Aires (UNCPBA-CICPBA-CONICET), University Campus, 7000 Tandil, Argentina.}

\affil[*]{abraham@uji.es}

\begin{abstract}
Optical matter, a rapidly growing field, aims to develop novel methods for controlling and assembling mesoscale systems through light-matter interactions. Traditional approaches often involve complex illumination fields with costly and unstable setups, requiring strong gradient forces, high-intensity laser spots that could harm samples, and substrate support. For binding, attractive inter-particle forces may not be sufficient to dynamically assemble systems due to unbalanced components such as centrifugal forces or collisions, leading to net repulsion between particles. In a previous work, magneto-optical nanoparticles illuminated with two counter-propagating circularly polarized waves were predicted to be optically bound under quasi-static conditions. However, the dynamics of such nanosystems were not thoroughly considered.
	
Here, a general framework to study magneto-optical (MO) matter is introduced. MO matter represents a radical concept for assembling small-scale systems through optical illumination, controllable by static magnetic fields. Dynamic binding between two n-doped InSb nanoparticles, which exhibit surface plasmons at THz frequencies, is recreated. Additionally, the reported examples may represent a novel approach to minimize possible sample damage using low-energy illumination sources without gradient components.
	
This numerical framework, grounded in Langevin dynamics and realistic collision phenomena, serves as a robust and universal tool for exploring various optomechanical designs where centrifugal and contact forces are present. The impact of thermal noise, collision types, initial conditions, and resonance excitation on a dimer system is examined, generating multiple scenarios for favorable dimer formation and subsequent manipulation. It is found that dimers tuned initially at magneto-optical resonance can bind in stable, average positions even when particle-particle collisions are present. The current method and subsequent findings are believed to provide essential knowledge for studying optical binding and the dynamics of any small-scale cluster of electromagnetically interacting sub-units.
\end{abstract}
\flushbottom
\maketitle
\thispagestyle{empty}

\section*{Introduction}

Concerning the top-down fabrication techniques, one of the most promising approaches relies on the use of optical forces (OF) to assemble clusters of mesoscale particles, since it allows one to manipulate a wide range of particles in different environments. No matter if these are active/living or passive \cite{ashkin_optical_1987,ashkin_internal_1989}, regular-shaped or irregular, electrically charged or neutral; OF can exert action on them in an non-invasive and safe, clean way. Every object is seen as a potential light absorber or scatterer since it is electromagnetically represented by  effective refractive index or dielectric/permeability functions. Also available at micro/nanoscale are the resonant properties of light-matter interaction like surface polaritons [e.g. phonons (SPhPs) and plasmons (SPPs)], Mie volume resonances, interband transitions and fluorescence \cite{brau_interlaced_2006,abraham_ekeroth_new_2018} (both of quantum nature), which can be found at relatively low-beam energies (from optical up to THz range). In several electromagnetic systems, the resonances can even be tuned and controlled by the use of simple external agents like magnetic fields. The use of magneto-optical (MO) materials can easily support this degree of resonance control. 

Tuning and controlling resonances are essential to enhance the photonic forces and make manipulation possible since other force types may appear in such small scales for example, the unavoidable thermodiffusive, drag (or viscous), thermophoretic \cite{kollipara_hypothermal_2023,zhou_low-temperature_2023,chand_emergence_2023}, acoustophoretic \cite{foresti_acoustophoretic_2018}, hydrodynamic \cite{herrera-velarde_hydrodynamic_2013,hong_stand-off_2020}, and electrophoretic forces \cite{zaman_modeling_2021}, among others. The study of stellar-dust particles being pushed by the sun's radiation \cite{weingartner_forces_2001,polimeno_optical_2021}, solar sails \cite{swartzlander_theory_2022}, optical levitation \cite{ashkin_optical_1975}, micro and nanosurgery \cite{blazquez-castro_genetic_2020}, microrheology \cite{volpe_roadmap_2023}, laser cooling \cite{castin_new_1989,arita_cooling_2023}, sorting \cite{hayat_lateral_2015}, information storing/processing \cite{he_towards_2022}, and manipulation of particles/living cells \cite{guck_stretching_2002,li_optical_2020,kishimoto_recent_2022}, constitute a few examples of application of opto-mechanics, whose basis is the electromagnetic conservation of momenta. 

The conservation of momentum implies that the complex, multiple scattering occurring at small scales may involve unbalanced torques as well as forces, which can be harnessed or utilized. For example, net torques are crucial in optical trapping and synchronized nanomotors or nanorotors \cite{la_porta_optical_2004,yan_three-dimensional_2012,cui_synchronization_2024}. Additionally, the photonic force microscopy became one of the most outstanding uses of the conservation laws of electromagnetic momenta \cite{lester_fundamentals_2001,nieto-vesperinas_optical_2015-1,pralle_photonic_2000}, and was proposed as an alternative to the atomic force microscopy since this latter requires macroscopic mechanical devices to guide the probe \cite{volpe_brownian_2007,magazzu_investigation_2023,crozier_plasmonic_2024}. In terms of angular momentum modification, the use of circularly-polarized illumination fields turned out to be essential, since they avoid the use of strongly focalized fields  \cite{cui_synchronization_2024}. However, in a dynamic scenario, unusual components like centrifugal forces may appear and play a key role, because they can turn the net forces repulsive. 
 
On the other hand, efficient separation or sorting methods have become essential for the development of nanodevices for high-demand manufacturing purposes. Connected with such degree of precision and knowledge about light-matter interaction is the concept of \textit{optical matter} (OM) \cite{burns_optical_1990,han_crossover_2018}, that have recently began to gain attention \cite{forbes_optical_2020,parker_optical_2020,cui_synchronization_2024}. Under certain conditions of coherence and synchronization,  moderated with an illuminating field, particles tend to self-organize and/or move dynamically as a whole entity. In this regard, a robust framework is required to understand the dynamic properties of such clusters, with the aim to predict and control the formation of \textit{photonic molecules} \cite{boriskina_photonic_2010}. Inherently related is the concept of optical trapping/tweezers, which uses carefully designed light spots to trap and move objects \cite{arbore_probing_2019,polimeno_optical_2021,crozier_plasmonic_2024}.   

Considerable amount of work have been done to comprehend and mimic the realistic conditions at which optical binding can occur to even explain several experimental outcomes  \cite{yan_three-dimensional_2012,sule_electrodynamics-langevin_2015,shang_assessing_2017,arita_coherent_2020}. In particular, models with high degree of detail like particle collisions have been included to understand microparticle trajectories that are induced by dielectrophoretic forces \cite{zaman_modeling_2021}. Furthermore, collisions could be characterized by means of dynamical imaging techniques in the case of fluorescent nanoparticles \cite{ma_dynamically_2018}, thus being useful for system characterization. However, all the approaches to particle impacts are usually constrained to elastic collisions where a complex richness of phenomena is expected due to plasticity \cite{millan_nucleation_2016}. This natural phenomenon has traditionally been underestimated and influences important mesoscale properties like heating, chemical reactions, and surface electric charging, for example in the performance of lithium batteries \cite{rano_characterization_2019} or other energy converters \cite{wang_light-controlled_2020}. 

On the downside, the typical OM setups involve sophisticated illumination fields, or structured light, that rely on gaussian counter-propagating beams, laguerre, bessel, or even vortex beams \cite{yevick_tractor_2016,nakajima_visualization_2022,he_towards_2022}, and need to be designed and held stable with high precision and accuracy, which make experiments less reliable and/or more expensive. This fact complicates even the analysis of single-particle motions \cite{nakajima_visualization_2022}. Besides, the usual optical traps require several milliwatts of light power that for optical or IR wavelengths could damage the sample \cite{volpe_roadmap_2023}. 

Following all these research gaps, the present work introduces a general framework to approach and understand \textit{magneto-optical (MO) matter}. MO matter is a novel concept to assemble small-scale systems by means of optical illumination and the influence of static magnetic fields on the coupled particles. Such a combination of effects offers a richer scenario for particle dynamics and optical binding \cite{edelstein_magneto-optical_2019,edelstein_magneto-optical_2021}. To keep the foundations simple, the study focuses on examples of the dynamics of dimers immersed in vacuum (or air) under magnetic field values of $0$ and $1$ T. It is found that particle collisions play a key role in the dynamics since strong attractive forces are expected by the MO effect. Centrifugal contributions are equally important due to the azimuthal symmetry imposed by the static magnetic field on the direction perpendicular to the initial dimer's geometry.

Specifically, the problem of optical binding between two coupled MO particles was addressed only very recently for n-doped InSb (n-InSb) \cite{abraham-ekeroth_numerical_2023}. n-InSb has been devised from InSb since it allows for sizable MO effects \cite{palik_coupled_1976,pandya_characterization_2015}. Nonetheless, the mechanical descriptions for such systems always approached "static" situations, where the full dynamics was not taken into account. For example, ref.~\cite{edelstein_magneto-optical_2021} reported equilibrium binding distances by tuning the polarization state and the static magnetic field. Notwithstanding, such a model is valid only for relatively low intensities of magnetic fields; moreover, it predicts stable dimers only when such magnetic field vector alternates, which contradicts the assumption of static. Ref.~\cite{abraham-ekeroth_numerical_2023} discussed possible rotations of the particles due to angular momentum transfer in the multiple scattering scheme, thus completing the mechanical predictions of the model but without taking full dynamics into account.

The illumination configuration is very simple and consists of two counter-propagating equal plane waves, both possessing the same circular polarization, as illustrated in Fig.~\ref{fig:1_GralConfig}. The model is based on a general Langevin formalism that deals with stochastic fluctuations in both linear and rotational degrees of freedom. It should be noted that research in optical manipulation and trapping in air/vacuum has shown renewed interest, which requires solutions for new technical challenges \cite{reimann_ghz_2018,shen_-chip_2021,volpe_roadmap_2023}. In particular, using a THz source for optical illumination can be a good strategy to reduce damage caused by heating. THz radiation typically has lower photon energy compared to visible or infrared light, which means it is less likely to cause thermal damage to sensitive samples \cite{salen_Matter_2019}. Additionally, THz sources can provide high spatial resolution and are useful for non-destructive imaging and spectroscopy \cite{salen_Matter_2019}. Examples of THz sources for optical manipulation can be found in \cite{mousavi_strong_2014,jia_efficient_2019,salen_Matter_2019,riccardi_electromagnetic_2023,huang_terahertz_2024}. Other ways to improve optical manipulation include producing samples sensitive to magnetic fields, altering the thermodynamic properties of the surrounding medium such as temperature and pressure, and functionalizing the particles to change their collision properties.

The present paper shows the possibility of building self-organized matter by extension of the case for a dimer of magneto-optical (MO) nanoparticles, which manifest binding phenomena by means of non-reciprocal forces \cite{edelstein_magneto-optical_2019,edelstein_magneto-optical_2021}. MO matter here is just considered as an introductory topic. Exhaustive exploration and optimization of the conditions to create OM such as polarization \cite{shi_dynamically_2023}, beam shapes, wavelengths, or consideration of several surrounding media is considered out of the scope of this paper.

\section*{Electromagnetic dimer interaction and induced mechanics.}

\begin{figure}
	\begin{centering}
		\includegraphics[width=8cm,keepaspectratio]{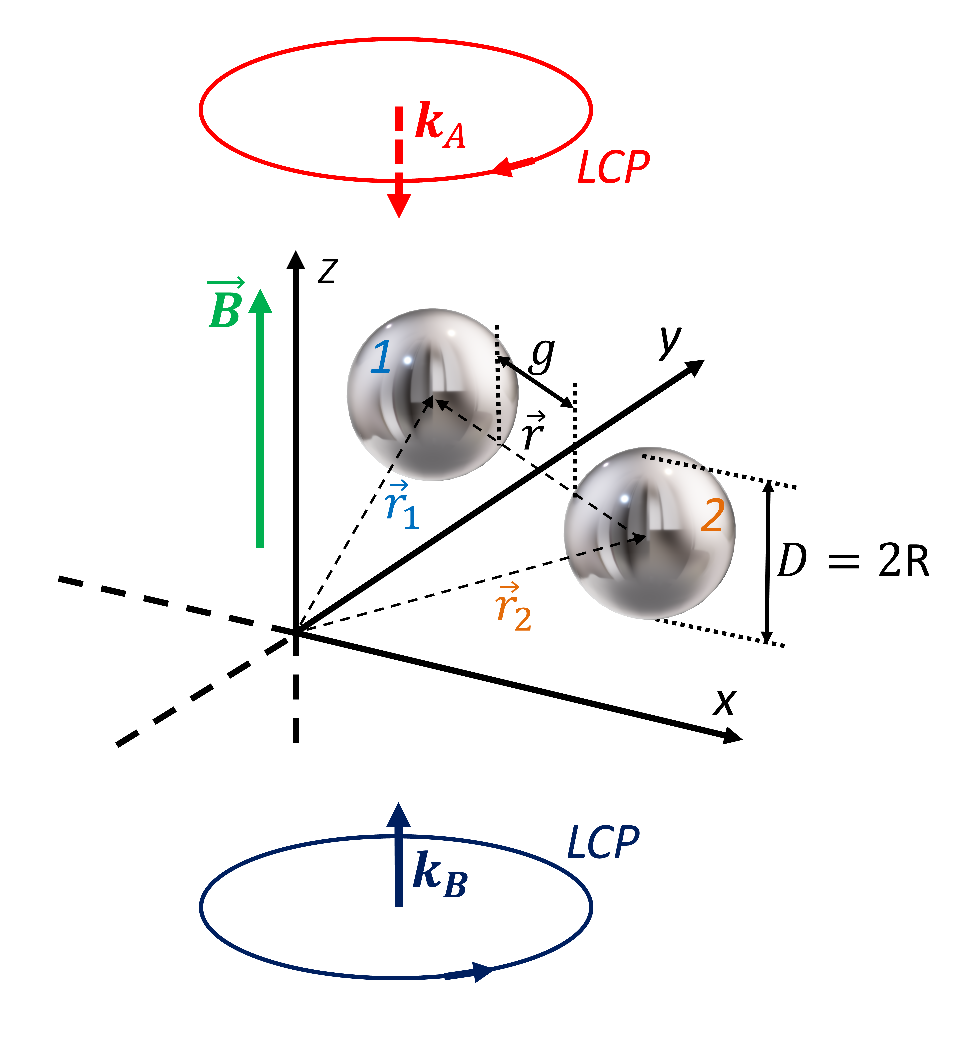}
		\par\end{centering}
	\caption{\label{fig:1_GralConfig} Geometric configuration of the problem. Any initial situation where the dimer's axis is parallel to the $(x,y)$ plane is called as "perpendicular alignment".}
\end{figure}

All the details about the constitutive and  optical properties of n-InSb particles simulated here can be found in  Ref.~\cite{palik_coupled_1976,abraham-ekeroth_numerical_2023}. The mechanical model for "static" dimers, namely without considering full dynamics, can also be found in Ref.~\cite{abraham-ekeroth_numerical_2023}. In particular, Refs.~\cite{abraham_ekeroth_thermal_2017,abraham_ekeroth_anisotropic_2018,edelstein_magneto-optical_2019,edelstein_magneto-optical_2021,edelstein_circular_2022,abraham-ekeroth_numerical_2023} explain how the magnetic field modifies the optical properties of these scatterers and breaks their symmetry. Here, only properties concerning dynamics are described. As the particles considered in this work are much smaller than the working wavelengths, they can be roughly modeled by their dipolar response. Let's avoid multipolar contributions that could be important for very short times when the particles are so close. The particles $1$ and $2$ are then represented by dipole moments $\mathbf p_{1}$-$\mathbf p_{2}$ respectively, which are coupled by the Green tensor $\hat G$ \cite{novotny_principles_2006} as
\begin{align} 
	\label{eq-dipoles-solved} 
	\mathbf p_{1} = \epsilon_0 \hat F \left(\mathbf E_{0,1} + k^2_0 \hat G \hat \alpha \mathbf E_{0,2}\right), \\ 
	\mathbf p_{2} = \epsilon_0 \hat F \left(\mathbf E_{0,2} + k^2_0 \hat G \hat \alpha \mathbf E_{0,1}\right). 
\end{align}
Here, $\epsilon_0$ is the vacuum permittivity, $k_0$ is the  wavenumber, $\mathbf E_{0,n}$ is the incident electric field at the dipoles' positions $\mathbf{r}_n$, $n=\{1,2\}$, and $\hat \alpha$ is the polarizability tensor that represents both particles, which includes the radiative corrections when it is expressed as follows \cite{ekeroth_optical_2019}:
\begin{equation}
	\label{eq-alpha}
	\hat \alpha = \left(\hat \alpha_0^{-1} - \frac{ik^3_0\hat I}{6\pi}\right)^{-1}
\end{equation}
where $\hat \alpha_0$ is the quasistatic polarizability, which for spherical dipolar particles can be given by
\begin{equation}
	\label{eq-alpha0}
	\hat \alpha_{0}^{-1} = \frac{1}{V} \left(\hat I/3 + [\hat \epsilon_r - \hat I]^{-1} 
	\right).
\end{equation}
where $V$ the particles' volume and $\hat \epsilon_r$ is the relative dielectric tensor. In Eq.~\ref{eq-dipoles-solved}, $\hat F= \left(\hat \alpha^{-1} - k^4_0 \hat G \hat \alpha \hat G\right)^{-1}$ in order to solve the coupled system. The illumination consists of a superposition of two counter-propagating, left-handed circularly polarized (LCP) plane waves with the same intensity $I_0$   \cite{cameron_optical_2014,edelstein_magneto-optical_2019}, see Fig.~\ref{fig:1_GralConfig}, namely,
\begin{equation}
	\label{eq-E0}
	\mathbf E_{0} = \frac{E_{0}}{\sqrt{2}}\left[\left(\mathbf{\check{x}}+i\mathbf{\check{y}}\right)e^{ik_0z}+\left(\mathbf{\check{x}}-i\mathbf{\check{y}}\right)e^{-ik_0z}\right].
\end{equation}
It is worth mentioning why this standing-wave field is chosen. This illumination easily creates a total field with zero-average values of $\nabla |\vec{E}|^2$, $\vec{S}$, and $\nabla \times \vec{J}_{spin}$, where $\vec{E}$, $\vec{S}$, and $\vec{J}_{spin}$ are the electric, Poynting, and spin density field. In other words, this total field induces zero net force on an isotropic particle in the absence of a static magnetic field $\vec{B}$,  \cite{edelstein_magneto-optical_2019}. Its total spin is always equal to zero, implying a null gradient force, while its helicity has a constant positive value, which implies a constant extinction force when $\vec{B}$ is on \cite{abraham-ekeroth_numerical_2023}. Besides avoiding complex optical traps, this illumination is homogeneous across space, which could help bind multiple particles to form large-scale optical matter, for example, using nanoparticles and laser illumination. 

The $i$-component of the forces exerted on each particle can be obtained from the time-averaged force within the Rayleigh approximation \cite{chaumet_time-averaged_2000}. This is 
\begin{eqnarray}
	\label{eq-FcDDA}
	F_{1,i}& = &\frac{1}{2}Re\{\mathbf{p}^t_1[\partial_i\mathbf{E}^{*}(\mathbf{r},\omega)|_{\mathbf{r}=\mathbf{r}_1}\} \\
	F_{2,i}& = &\frac{1}{2}Re\{\mathbf{p}^t_2[\partial_i\mathbf{E}^{*}(\mathbf{r},\omega)|_{\mathbf{r}=\mathbf{r}_2}\}
\end{eqnarray}
where the derivatives of the total field $\partial_i\mathbf{E}(\mathbf{r},\omega)|_{\mathbf{r}=\mathbf{r}_n}$ at the dipoles' positions $\mathbf{r}_n$ can be obtained from \cite{chaumet_coupled_2007}:
\begin{align}
	\label{eq-derivsEr}
	& \partial_i\mathbf{E}(\mathbf{r},\omega)|_{\mathbf{r}=\mathbf{r}_1}=\partial_i\mathbf{E}_0(\mathbf{r},\omega)|_{\mathbf{r}=\mathbf{r}_1}+ \nonumber \\ 
	& +\frac{k^2_0}{\epsilon_0}(\partial_i\mathbf{G}(\mathbf{r},\mathbf{r}_2))_{\mathbf{r}=\mathbf{r}_1}\mathbf{p}_2]\}  \\
	&\partial_i\mathbf{E}(\mathbf{r},\omega)|_{\mathbf{r}=\mathbf{r}_2}=\partial_i\mathbf{E}_0(\mathbf{r},\omega)|_{\mathbf{r}=\mathbf{r}_2} + \nonumber \\ 
	& +\frac{k^2_0}{\epsilon_0}(\partial_i\mathbf{G}(\mathbf{r}_1,\mathbf{r}))_{\mathbf{r}=\mathbf{r}_2}\mathbf{p}_1]\} 
\end{align}
The total force exerted on the dimer results from adding the force components for each particle, namely, $F_{tot,i} = F_{1,i}+F_{2,i}$. In particular, the net radiation pressure for the dimer under the illumination given by Eq.~\ref{eq-E0} is defined by taking $i=3$, or the $z$ components, as
\begin{equation}
	\label{eq-ESys}
	F_{tot,z} = F_{1,z}+F_{2,z}
\end{equation}
Another useful mechanical variable is the binding force, which in the present case is defined as
\begin{equation}
	\label{eq-BindingF}
	\Delta = \left(\mathbf{F}_1 - \mathbf{F}_2\right)\cdot \check{n}
\end{equation}
where $\mathbf{\check{n}}=\frac{\mathbf{r}_2-\mathbf{r}_1}{|\mathbf{r}_2-\mathbf{r}_1|}$ is the dimer's versor. The optical torques can also be calculated, as given in Ref.~\cite{chaumet_electromagnetic_2009}:
\begin{align}
	\label{eq-Tqs1}
	& \mathbf{N}_{spin,1} = \frac{1}{2\epsilon_0}Re\left\{\mathbf{p}_{1} \times \left[\left(\hat \alpha_0^{-1}\right)^*\mathbf{p}_1^*\right]\right\} \\
	& \mathbf{N}_{orb,1} = \mathbf{r}_{1} \times \mathbf{F}_{1} \\
	& \mathbf{N}_{1} = \mathbf{N}_{spin,1} + \mathbf{N}_{orb,1}
\end{align}
\begin{align}
	\label{eq-Tqs2}
	& \mathbf{N}_{spin,2} = \frac{1}{2\epsilon_0}Re\left\{\mathbf{p}_{2} \times \left[\left(\hat \alpha_0^{-1}\right)^*\mathbf{p}_2^*\right]\right\} \\
	& \mathbf{N}_{orb,2} = \mathbf{r}_{2} \times \mathbf{F}_{2} \\
	& \mathbf{N}_{2} = \mathbf{N}_{spin,2} + \mathbf{N}_{orb,2}
\end{align}
The definitions of the orbital and spin torques were discussed previously in Refs.~\cite{nieto-vesperinas_optical_2015,nieto-vesperinas_optical_2015-1,chaumet_electromagnetic_2009}, among others. The spin torques are always defined with respect to the centers of the particles. Otherwise, the reference system is located at the dimer's center of mass, and orbital torques are set. 

\section*{Linear Langevin equation and its discretization \label{sec:LinearLangevin}}

We deduce a method similar to that reported in Ref.~\cite{zaman_modeling_2021} for an easy integration of the collision physics into the problem. However, with the spirit of unifying notation among the literature and later introducing novel contributions, let's briefly recapitulate such deduction. Assume a particle of mass $m$ and size $2R$ moving with velocity $\vec{v}$ in a shear flow with low Reynolds number. The particle's center of mass (CM) is located at position $\vec{r}$ in time $t$
\begin{equation}
m \frac{\partial \vec{v}(\vec{r},t)}{\partial t} = \lambda_\eta \left[\vec{v}_f(\vec{r},t) - \vec{v}(\vec{r},t) \right] + \vec{F}_{ext}(\vec{r},t) + \vec{W}(t), \label{eq:LangevinF}
\end{equation}
such that $\vec{v}_f$ is the velocity of the fluid or embedding medium, $\lambda_\eta$ is the friction coefficient given by Stokes' law, namely, $\lambda_\eta = 6\pi \eta R$, where $\eta$ is the dynamic viscosity of the medium. $\eta$ may depend on the temperature $T$ through different models. For example, an empirical model might evaluate \cite{reif_fundamentals_2009}
\begin{equation}
		\eta=2.791\times 10^{-7}T^{0.7355} \label{eq:eta_funct_T}
\end{equation}
which gives $\eta$ in $Pa.s$ when $T$ is in K, and is accurate between (-20,400) $^\circ C$. 
At very low temperatures, the Stokes' law can still be used as an approximation if corrected as $\lambda_\eta \rightarrow \lambda_\eta/C$, where $C=1 + l/R(A_1 + A_2exp(-A_32R/l))$ is the Cunningham correction factor, with $A_j$ being known constants \cite{cunningham_velocity_1997,davies_definitive_1945}. Furthermore, to avoid dealing with rarefied gas phases and keep using the \textit{continuous} Brownian formulation for the surrounding medium, the Knudsen number $K_n=\frac{l}{2R}$ must be kept low, which means to constrain the model to low (molecular) mean free paths, $l$, compared to the size $2R$ \cite{reif_fundamentals_2009}. In eq.~\ref{eq:LangevinF}, $\vec{F}_{ext}$ is the external force (e.g. optical) and $\vec{W}(t)$ is a term for the stochastic thermal noise with a Gaussian probability distribution. The fluctuation-dissipation theorem (FDT) connects $\lambda_\eta$ and $\vec{W}(t)$ by the result that
\begin{equation}
\left\langle \vec{W}(t) \otimes \vec{W}(t') \right\rangle = 2\lambda_\eta k_B T \hat 1 \delta (t-t') \label{eq:FDT_forces}
\end{equation}
where $\hat 1$ is the unit tensor, and $\left\langle\vec{W}(t)\right\rangle=\vec{0}$. Equation~\ref{eq:LangevinF} describes the motion of the particle under the influence of both the drag and the Brownian forces at temperature $T$. Eq.~\ref{eq:LangevinF} can be put another way by using the so-called diffusion tensor $\hat D$. Assuming an isotropic and homogeneous embedding medium, this tensor can be simplified to $\hat D=D_0 \hat 1$, where $D_0=\frac{k_B T}{\lambda_\eta}=\frac{k_B T}{ 6\pi \eta R}$. Note that $\vec{W}(t)$ has units of force; however, it can be expressed by a dimensionless white-noise vector $\vec{w}$ that holds its properties of zero mean and unit variance \cite{sule_electrodynamics-langevin_2015}. Thus, $\vec{W}(t)=\sqrt{m}\sigma \sqrt{\frac{d\vec{w}}{dt}}$, where $\sigma = \sqrt{2 \gamma k_B T}$ and $\gamma=\frac{\lambda_\eta}{m}$ represents the strength of the random forces \cite{shang_assessing_2017}. $\sqrt{\frac{d\vec{w}}{dt}}$ should be interpreted as the element-wise square root of the vector. An alternative formulation commonly used in BAOBAB methods uses a three-dimensional random vector $\vec{\mathcal{R}}(t)$ with dimensions of $1/t^{1/2}$ \cite{sule_electrodynamics-langevin_2015}, which is linked to $\vec{W}(t)$ by $\vec{W}(t)=\sqrt{2}\sqrt{\lambda_\eta k_B T}\vec{\mathcal{R}}(t)$. In this way, Eq.~\ref{eq:LangevinF} can be expressed as
\begin{align}
	m \frac{\partial \vec{v}(\vec{r},t)}{\partial t} &= k_B T \hat D^{-1}
	\lambda_\eta \left[\vec{v}_f(\vec{r},t) - \vec{v}(\vec{r},t) + \sqrt{2} \hat D_\frac{1}{2}\vec{\mathcal{R}}(t) \right] \nonumber \\
	+ \vec{F}_{ext}(\vec{r},t), \label{eq:LangevinFD}
\end{align}
where $\hat D_\frac{1}{2}$  is defined as the element-wise square root of $\hat D$ \cite{zaman_modeling_2021}. To discretize the problem, it is convenient to use the approximations $\frac{\partial \vec{v}}{\partial t} \approx \frac{\vec{v}_{n+1} - \vec{v}_{n}}{\Delta t}$, $\vec{\mathcal{R}}(t) \approx \frac{\vec{R_n}}{\sqrt{\Delta t}}$, where $\vec{R_n}=\sqrt{\Delta \vec{W}(t)}$ is a dimensionless source of fluctuations. Thus, the discretized version of Langevin equation~\ref{eq:LangevinFD} is
\begin{align}
	m \frac{\vec{v}_{n+1} - \vec{v}_{n}}{\Delta t} &= k_B T \hat D^{-1}
	\lambda_\eta \left[\vec{v}_{f,n} - \vec{v}_{n+1} + \sqrt{\frac{2}{\Delta t}} \hat D_\frac{1}{2}\vec{R}_n \right] \nonumber \\
	+ \vec{F}_{ext,n}, \label{eq:LangevinFDDisc}
\end{align}
where $\vec{r}_n = \vec{r}(t_n)$ is the position of the particle's CM at time $t_n$ of the simulation, and $\vec{F}_{ext,n}\equiv \vec{F}_{ext}(\vec{r}_n)$. $\Delta t = \frac{t_f - t_0}{N-1}$ if $t_0, \, t_f$ are the initial and final time values to evaluate, and $N$ is the number of time values. Note that $\Delta t \rightarrow dt$ if $N \rightarrow \infty$.

As eq.~\ref{eq:LangevinFDDisc} is a numerical basis for the solution of a stochastic differential equation, this must be stabilized to obtain realistic results. By performing algebraic operations on eq.~\ref{eq:LangevinFDDisc}, the velocity at the next step of the simulation can be obtained by all the magnitudes at the current step, as
\begin{align}
	\vec{v}_{n+1} &= \hat \Lambda \left(\hat \Lambda + \hat 1 \right)^{-1} \vec{v}_n + \nonumber \\
	&+ \left(\hat \Lambda + \hat 1 \right)^{-1}\left[\vec{v}_{f,n} + \sqrt{\frac{2}{\Delta t}} \hat D_\frac{1}{2}\vec{R}_n + \frac{\hat D}{k_B T}\vec{F}_{ext,n}\right], \label{eq:LangevinFDFinal}
\end{align}
where $\hat \Lambda = \frac{m D_0}{k_B T \Delta t} \hat 1$.

\section*{Rotational Langevin equation and its discretization}

A similar formulation to that given in Sec.~\ref{sec:LinearLangevin} can be applied to the problem of the particle rotation. In this case, let's consider a particle inmmersed in a flow which could be rotating at angular velocity $\omega_f$. The particle can also rotate at angular velocity $\omega$ around its CM, under the influence of an external torque (i.e., optical, magnetic, electrophoretic, etc.) and/or a torque coming from the stochastic thermal noise. The Langevin corresponding equation is then,
\begin{equation}
	\hat{I} \frac{\partial \vec{\omega}(\vec{r},t)}{\partial t} = \lambda_{R,\eta} \left[\vec{\omega}_f(\vec{r},t) - \vec{\omega}(\vec{r},t) \right] + \vec{\tau}_{ext}(\vec{r},t) +\vec{\Upsilon}(t), \label{eq:LangevinR}
\end{equation}
here $\hat{I}$ is the moment of inertia of the particle and $\lambda_{R,\eta}=8 \pi \eta R^3/C$ is the rotational friction coefficient, where $C$ is the above defined Cunningham factor.  FDT can be also applied to the thermal torque
\begin{equation}
	\left\langle \vec{\Upsilon}(t) \otimes \vec{\Upsilon}(t') \right\rangle = 2\lambda_{R,\eta} k_B T \hat 1 \delta (t-t')  \label{eq:FDT_torques}
\end{equation}
and its first average is $\left\langle\vec{\Upsilon}(t)\right\rangle=0$. So analogously to what was done for the diffusion tensor, $\hat E = E_0 \hat 1$ can be defined for an isotropic medium, where $E_0 = \frac{k_B T}{\lambda_{R\eta}}= \frac{k_B T}{8 \pi \eta R^3}$ such that it has dimensions of $1/t$. By performing a dimensional analysis, $\vec{\Upsilon}(t)=\sigma_R \hat{I}^{\frac{1}{2}} \sqrt{\frac{d\vec{T}}{dt}}$ can be defined, where the element-wise operation for root squaring was again used (do not confuse the vector $\vec{T}$ with the temperature $T$). Then $\sigma_R=\sqrt{2 \gamma_R k_B T}$ is a novel variable to mean strength of the random torques; concerning particles made of isotropic materials, $\hat{I}=I_0 \hat 1$ so that  $\gamma_R=\frac{\lambda_{R\eta}}{I_0}$ can be defined. For a symmetric formulation, $\Upsilon$ can be also expressed as a function of the rotational equivalent of the vector $\vec{\mathcal{R}}(t)$, as $\Upsilon(t)=\sqrt{2 \lambda_{R\eta}k_B T}\vec{\mathcal{S}} (t)$. In total correspondence, $\vec{\mathcal{S}}(t)$ has dimensions of $1/t^{1/2}$ and provides the brownian variation for the torques. It allows us to write Eq.~ \ref{eq:LangevinR} as
\begin{align}
	\hat I \frac{\partial \vec{\omega}(\vec{r},t)}{\partial t} &= k_B T \hat E^{-1}
	\lambda_\eta \left[\vec{\omega}_f(\vec{r},t) - \vec{\omega}(\vec{r},t) + \sqrt{2} \hat E_\frac{1}{2}\vec{\mathcal{S}}(t) \right] \nonumber \\
	+ \vec{\tau}_{ext}(\vec{r},t), \label{eq:LangevinTE}
\end{align}
To discretize this equation, consider $\frac{\partial \vec{\omega}}{\partial t} \approx \frac{\vec{\omega}_{n+1} - \vec{\omega}_{n}}{\Delta t}$, $\vec{\mathcal{S}}(t) \approx \frac{\vec{S_n}}{\sqrt{\Delta t}}$, where $\vec{S_n}=\sqrt{\Delta \vec{T}(t)}$ is a dimensionless source of fluctuations. Following the same steps as in the previous section to obtain eq.~\ref{eq:LangevinFDFinal}, one can get for the angular velocity
\begin{align}
	\vec{\omega}_{n+1} &= \left(\hat{\Lambda}_R + \hat 1 \right)^{-1}\hat{\Lambda}_R \vec{\omega}_n + \nonumber \\
	&+ \left(\hat{\Lambda}_R + \hat 1 \right)^{-1}\left[\vec{\omega}_{f,n} + \sqrt{\frac{2}{\Delta t}} \hat E_\frac{1}{2}\vec{S}_n + \frac{\hat E}{k_B T}\vec{\tau}_{ext,n}\right], \label{eq:LangevinTEFinal}
\end{align}
where $\hat{\Lambda}_R = \frac{\hat E \hat I}{k_B T \Delta t}$.
Noteworthy, both Langevin formulations for forces and torques can be applied to any number of particles entering the problem, e.g., dimers. The movements will be coupled by the nature of the external forces (like in optical binding) and by the collision processes occurring between the interacting particles. Because of this, it is important to remark that there is no need to deal with explictly coupled Langevin equations like those in Ref.~\cite{volpe_torque_2006}; the coupling in the present dynamics is implicit. The random vectors are generated by the Box-Mueller method with additional normal distribution generators \cite{sule_electrodynamics-langevin_2015}.

Subsequently, all the different components of the mechanical magnitudes can be obtained from this formulation. Here, the relevant variables are written below for convenience
\begin{align}
\vec{F}_{drag} &= \lambda_\eta \left(\vec{v}_f - \vec{v} \right), \\
\vec{W}(t) &= \sqrt{2 \lambda_\eta k_B T} \vec{\mathcal{R}}, \\
\vec{\tau}_{drag} &= \lambda_{\eta R} \left(\vec{\omega}_f - \vec{\omega} \right), \\
\vec{\Upsilon}(t) &= \sqrt{2 \lambda_{\eta R} k_B T}\vec{\mathcal{S}}(t).
\end{align}
\section*{Formulation for Particle-particle collisions.}

\begin{figure}
	\begin{centering}
		\includegraphics[width=8cm,keepaspectratio]{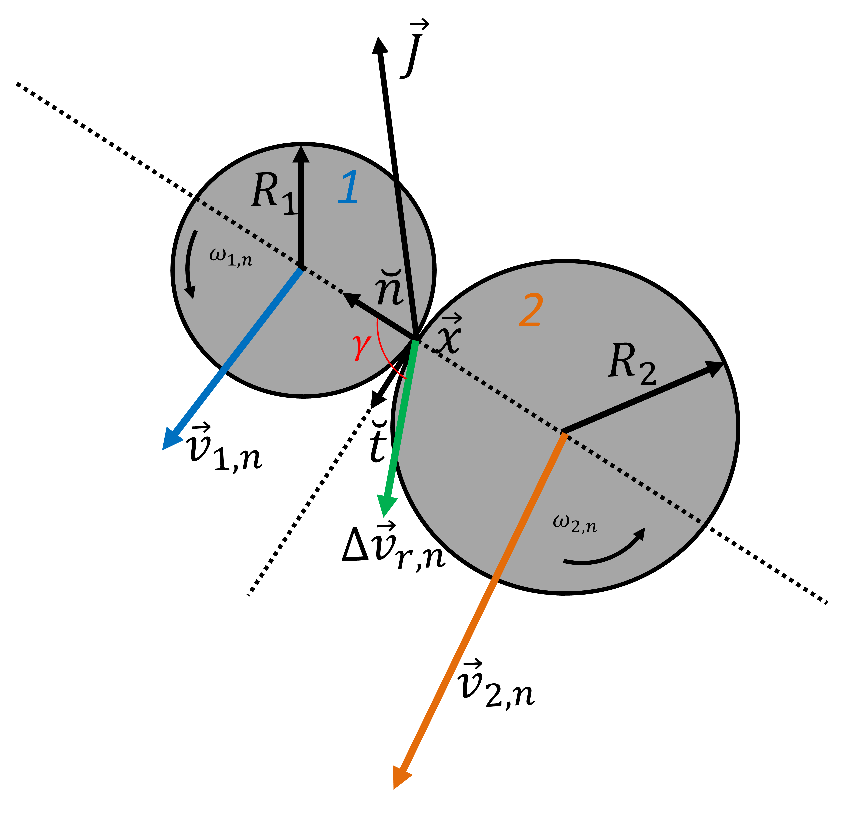}
		\par\end{centering}
	\caption{\label{fig:ConfigCollisions} Diagram of the dimer just before collision at the current step, $n$, of the simulation. $\vec{X}$ is the contact point between the sphere in the moment of collision.}
\end{figure}

In this section, the collision physics is applied to model the particles as two rigid bodies interacting by friction forces. The model is stated for a dimer but it can be easily extended to many particles as desired. By applying a relatively simple formulation \cite{orlando_effect_2010}, one can obtain the updated velocities after collision, the impulse interchanged, and the friction coefficient if a few initial parameters like the mass and the inertia tensor are known. The scheme in Fig.~\ref{fig:ConfigCollisions} helps with the definitions used in this section. Following the discretization method chosen and based on the definition of the impulse 
\begin{align}
	\vec{J}_n &= \int_{t_n}^{t_{n+1}}\frac{\Delta \vec{P}}{\Delta t}dt \approx \Delta \vec{P} \Bigr|_{t_n}^{t_{n+1}} = \Delta \vec{P}_n \label{eq:impulse_Jn}
\end{align}
From definition \ref{eq:impulse_Jn} and using the Newton's third law, we get the linear momentum interchanged between the spheres
\begin{align}
	\Delta \vec{P}_{1,n} &= m_1 \left(\vec{v}_{1,n+1} - \vec{v}_{1,n} \right) = \vec{J}_n, \label{eq:DP1} \\
	\Delta \vec{P}_{2,n} &= m_2 \left(\vec{v}_{2,n+1} - \vec{v}_{2,n} \right) = -\vec{J}_n. \label{eq:DP2} \\
\end{align}
As the impulses $\pm \vec{J}_n$ are just internal and external forces are negligible for $\Delta t_n \rightarrow 0$, the whole linear momentum is conserved $\Delta \vec{P}^{syst}_{n} = \Delta \vec{P}_{1,n} + \Delta \vec{P}_{2,n}=0$. Analogously, for angular momenta,
\begin{align}
	\Delta \vec{L}_{1,n} &= \hat{I}_1 \left(\vec{\omega}_{1,n+1} - \vec{\omega}_{1,n} \right) = \vec{R}_1 \times \vec{J}_n - \vec{K}, \label{eq:DL1} \\
	\Delta \vec{L}_{2,n} &= \hat{I}_2 \left(\vec{\omega}_{2,n+1} - \vec{\omega}_{2,n} \right) = \vec{R}_2 \times \left(-\vec{J}_n\right) + \vec{K}, \label{eq:DL2}
\end{align}
such that $\Delta \vec{L}^{syst}_{n} = \Delta \vec{L}_{1,n} + \Delta \vec{L}_{2,n}=0$, where $\vec{R}_1 = -R_1\breve n$ and $\vec{R}_2 = R_2\breve n$ (see Fig.~\ref{fig:ConfigCollisions}). The velocities at the contact point are useful, 
\begin{align}
	\vec{v}_{1,n}(\vec{X}) &= \vec{v}_{1,n} + \omega_{1,n} \times \vec{R}_1 \label{eq:v1n_X}, \\
	\vec{v}_{2,n}(\vec{X}) &= \vec{v}_{2,n} + \omega_{2,n} \times \vec{R}_2 \label{eq:v2n_X}, \\
\end{align}
so that the relative velocity difference at the contact point is $\Delta \vec{v}_{r,n}(\vec{X})=\vec{v}_{1,n}(\vec{X})-\vec{v}_{2,n}(\vec{X})$. Presently, the coefficient of normal restitution can be defined by the relation between the relative velocity differences after and before the collision, namely, $\breve n \cdot \vec{v}_{r,n+1}(\vec{X})= -e \breve n \cdot \vec{v}_{r,n}(\vec{X})$. Note that $0 \leq e \leq 1$. By using this definition together with eqs.~\ref{eq:DP1}-\ref{eq:DP2} and \ref{eq:v1n_X}-\ref{eq:v2n_X}, we can leave an expression for the impulse that does not depend on the contact point nor the next step in the simulations, i.e.,
\begin{align}
	\vec{J}_n &= -m^* (1+e)\left[\vec{v}_{1,n} - \vec{v}_{2,n} - \left(R_1\omega_{1,n} + R_2\omega_{2,n}\right) \times \breve n \right], \label{eq:Jn}
\end{align}
where $m^*$ is the reduced mass of the system. Now all the characteristic variables of the movement can be updated at step $n+1$ by using the impulse value at step $n$. The velocities can be cleared from eqs.~\ref{eq:DP1}-\ref{eq:DL2}, and all the linear and angular positions of the particles can be updated by using the Euler-Cromer method \cite{cromer_stable_1981}. The accelerations are recovered by employing the time interval to divide the velocities at two succesive steps. In addition, the coefficient of sliding friction $\mu$ can also be estimated from the classical definition for the magnitude of the sliding friction force
\begin{align}
	\mu_n &= \frac{\left|\breve n \times \vec{J}_n \right| }{\left|\breve n  \cdot \vec{J}_n \right| }= \frac{\left|\vec{J}_n \cdot \breve t\right| }{\left|\vec{J}_n \cdot \breve n\right| }. \label{eq:SlidingFriction}
\end{align}
Is it important to remark that a special algorithm had to be designed to check the collisions between the particles since these events depend numerically on the minimum gap considered to distinguish two touching objects as well as the time step on each simulation stage $n$. In this regard, such algorithm was prepared by checking collisions like that in Ref.~\cite{zaman_modeling_2021} but ensuring the correspondence in both linear and angular variables. This code was built into MATLAB\textsuperscript{\tiny\textregistered} R2024a and run in a computer with an Intel Core i7-8700 3.2 GHz processor, 32GB of DDR4 RAM, and a 1TB SSD.

\section*{Results and Discussion}

For small particles compared to the wavelength, the n-InSb dimer can be outstandingly represented by two coupled dipoles \cite{abraham_ekeroth_thermal_2017,abraham_ekeroth_anisotropic_2018,edelstein_magneto-optical_2019,abraham-ekeroth_numerical_2023}. In particular, given two equal particles of radius $R=200$ nm, it is already well-known that the resonances will be tunable by the dimer's gap $g$ and the static magnetic-field vector $B$, whether present. Following the indications from a recent study \cite{abraham-ekeroth_numerical_2023}, the minimal value for the initial gap in the dynamic simulations is $g_0 = 0.1R$. For a brief report, the formulation is constrained to static air environment as surrounding medium, such that $\vec{v}_{f,n} = 0$ and $\vec{\omega}_{f,n} = 0$ in Eqs.~\ref{eq:LangevinFDFinal} and \ref{eq:LangevinTEFinal} respectively for all $n$. All the results shown below correspond to initial  "perpendicular" alignment of the dimer with respect to the illuminating beams, namely, the dimer's axis lie perpendicular to the optical axis of the illumination. Initial "Parallel" configurations were also studied, but all lead to repulsive states and thus dimer's dissolution, in agreement with the "static" results presented in Ref.~\cite{abraham-ekeroth_numerical_2023}. Though the method is fully general, all the studies are carried on with zero initial velocities for succinctness. Following the common literature, all the simulations here last for a few ms (up to 10 ms in the present work). The values simulated correspond to $I_0=4 \, mW/\mu m^2$, which is a typical value of illumination intensity \cite{edelstein_magneto-optical_2019,crozier_plasmonic_2024}.

In principle, it would be necessary to account for van der Waals (vdW) interactions when the particles become close to each other. Such forces are significant when the relative sizes are small, i.e., interparticle gaps and/or particle diameters, in the order of a few nanometers \cite{hamaker_londonvan_1937,novotny_van_2008,yannopapas_first-principles_2007}. A rapid estimation of the order of magnitude of these forces can be done with the help of Hamaker's formulation for Lifshitz-Van der Waal forces between spheres of equal radii $R$ and gap $g$, 
\begin{equation}
	F_{vdW}=-\frac{AR}{12g^2}, \label{eq:Hamaker-vdW}
\end{equation}
where $A$ is the Hamaker coefficient, which is a constant ($\sim 10^{-19} - 10^{-20}$ Joules) that depends on the materials involved. To obtain a rough number $\mathcal{O} \left(|F_{vdW}|\right)$, let’s consider that $\mathcal{O}(A)\leq 10^{-19}$ Joules and $g=g_0$ for the minimal gap at the initial stage of the movement. Hence, $\mathcal{O}\left(|F_{vdW}|\right)=4.17$ pN is obtained for radius $200$ nm. Note that this number is of the same order as or smaller than the MO forces induced in the dimer only when no collisions are present. If collisions are included, the contact forces will generally be much larger than Van der Waal forces. For extremely short values of $g$, similar calculations by Ref. \cite{yannopapas_first-principles_2007} for a silver dimer of small particles ($R \sim 10$ nm) at $T=0$ K give $\mathcal{O}\left(|F_{vdW}|\right)\sim 100$ pN when $g<1$ nm, claiming that the $F_{vdW}$ indeed does not follow the $g^{-2}$ law for $g \rightarrow 0$ but decreases exponentially for values $0<g<3$ nm. For particles at higher temperatures, it is expected that the vdW forces are reduced by about ten times the values for $T=0$ K \cite{henkel_radiation_2002}. Overall, including vdW forces is considered beyond the scope of this work since it does not significantly contribute to the current purposes.

\subsection*{Quasi-determinist behavior at temperature near the absolute zero: dynamics with negligible Brownian fluctuations}

To first comprehend the dynamics of the MO dimers, the study focuses on an \textit{extreme} example for very low-temperature, namely, $T = 5$ mK, where the thermal noise can be practically neglected. This limit can help understand problems of molecular or mesoscale dynamics involving any other sources of forces and illuminations, as for instance those reported in \cite{hong_stand-off_2020,zhou_low-temperature_2023,kollipara_hypothermal_2023}. To validate the results, the Knudsen number was estimated by using ideal gas theory, as $K_n = \frac{k_bT}{\sqrt{2}\pi d^2 P_0}$, where $P_0$ is the working pressure and $d$ is the molecular diameter, which for air is about $0.37$ nm. A value $P_0>0.2837$ mbar, such that $K_n<0.01$, ensures a continuum fluid approximation for air. In addition, a rough extrapolation of $\eta$ to $\eta(5 \, mK)$ using eq.4~\ref{eq:eta_funct_T} allows us to assess the MO forces exerted on the dimer as a first attempt to comprehend MO matter. 

\begin{figure}[h]
	\begin{centering}
		\includegraphics[width=9cm,keepaspectratio]{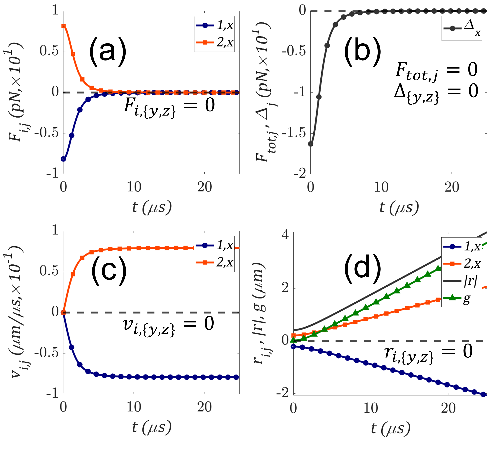}
		\par\end{centering}
	\caption{\label{fig:Dyn_T5mK_B0T_wl48p56um} Dynamics for the dimer when n-InSb is isotropic, $B = 0$, during the first moments at a temperature of $T_0 = 5$ mK. The dimer is initially tuned to its SPhP at $\lambda = 48.56 \mu$m for an initial gap of $g_0 = 0.1R$ and perpendicular alignment. (a) Optical forces. (b) Total forces exerted on the system. (c) Velocities. (d) Particle positions,  reduced-mass coordinate, and gap as a function of time. The dash lines correspond to zero and highlight the response by the components labeled on each panel. Noise forces are negligible in this limit, i.e., $T_0 \rightarrow 0$.}
\end{figure}

When the static magnetic field is off ($B=0$) and $g=0.1R$\,(=const.), the dimer's absorption resonances are located at $48.56 \mu$m (SPhP) and $72.384 \mu$m (SPP, not shown) \cite{abraham-ekeroth_numerical_2023}.  Fig.~\ref{fig:Dyn_T5mK_B0T_wl48p56um} shows results for this dimer when the initial gap is $g_0=0.1R$ and the illumination wavelength is set at  $\lambda = 48.56 \mu$m, which is the strongest response of the spectrum.
By studying the curves for the dynamics in Fig.~\ref{fig:Dyn_T5mK_B0T_wl48p56um}, it becomes quite clear that the configuration leads to a repulsive state, in which the dimer cannot be formed and no particle collisions are detected. Panel (a) shows the optical forces exerted on each particle; (b) the net forces on the system, including the radiation pressure, or forces exerted on the center of mass (CM) of the system ($F_{tot,j}$) ($j=x,y,z$), and the binding components $\Delta_j$; (c) the velocities for each particle; and (d) the particle positions, including the evolution of the gap $g$, and the reduced-mass coordinate $r$ with time. For clarity, only the first instants of the simulations are presented. The high-symmetry along $z$ is also evident from these curves when $x,y$, and $z$ components are compared.

Below, we will see that switching on/off the magnetic field constitutes a simple and powerful method to control the dimer's creation and movement. In fact, the movements do not turn out so sensitive to the resonance wavelength excited since that only matters for the very initial instants, so relaxing that kind of precision for the illumination makes the results more robust. This lack of sensibility with the resonance wavelength is also valid for dimers made of anisotropic n-InSb (not shown). 
\begin{figure}[h]
	\begin{centering}
		\includegraphics[width=9cm,keepaspectratio]{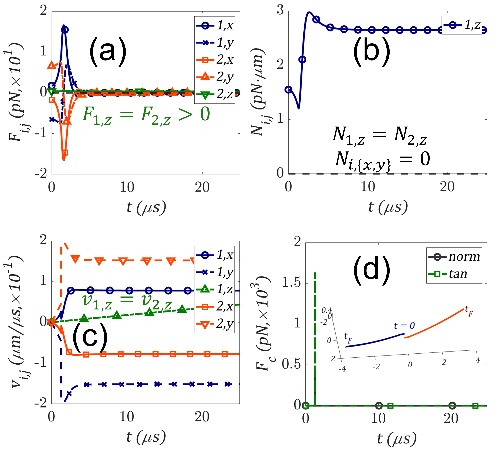}
		\par\end{centering}
	\caption{\label{fig:Dyn_T5mK_B1T_e1_wl49p81um} Dynamics for the dimer made of (anisotropic) n-InSb during the first moments, i.e. $B = 1$ T at $T_0 = 5$ mK. The dimer is initially tuned to one of its SPhPs at $\lambda = 49.81 \mu$m for an initial gap of $g_0 = 0.1R$ and perpendicular alignment. (a) Optical forces. (b) Optical torques. (c) Velocities. (d) Contact forces and perspective of the particle trajectories (inset). Noise forces are negligible in this limit, i.e., $T_0 \rightarrow 0$.}
\end{figure}
Fig.~\ref{fig:Dyn_T5mK_B1T_e1_wl49p81um} presents the situation of the dimer when $B = 1$ T, which also deals with a repulsion state due to particle-particle collision. The wavelength $\lambda = 49.81 \mu$m tunes one SPhP that results from the MO splitting of the SPhP located at $48.56µm$ (dimer's resonance for isotropic n-InSb)  \cite{abraham_ekeroth_anisotropic_2018}. Panel (a) [b] shows the induced force [torque] components on each particle; (c) shows the resultant velocities, and (d) the magnitudes of the contact forces because one collision has occurred in the process. The inset displays the particles' trajectories. The collision event can be observed in all the panels; a sudden change occurs in all the magnitudes. In particular, the contact forces are so strong that lead to the particles' repulsion. Both panels (b) and (d) manifest the presence of the rotational dynamics; specifically, the tangential forces in (d) present a higher peak than normal's. Rotational dynamics also appears for the isotropic case of Fig.~\ref{fig:Dyn_T5mK_B0T_wl48p56um}; however, the effects are negligible (not shown for brevity).

Remarkably, the initial conditions of the movement play a key role on the dynamics so that other more complex, helical trajectories can be obtained by properly varying the initial gap of the problem. Helical trajectories are expected along $z$ due to the inherent azimuthal symmetry imposed by the magnetic field's presence.
\begin{figure}[h]
	\begin{centering}
		\includegraphics[width=9cm,keepaspectratio]{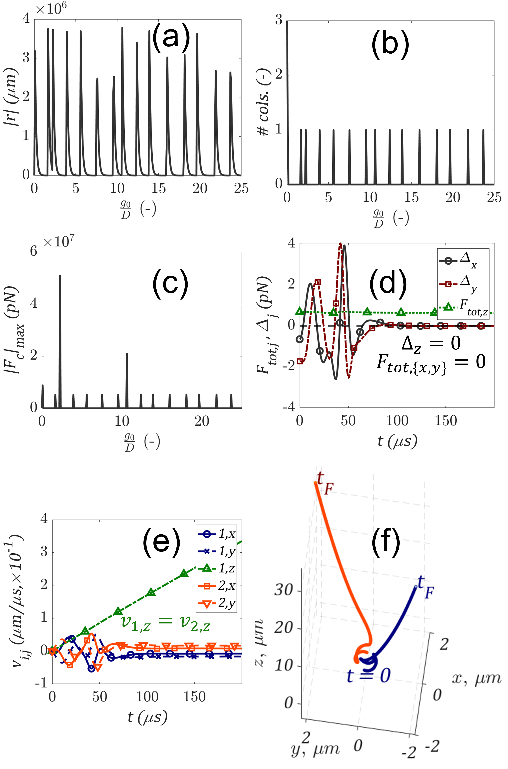}
		\par\end{centering}
	\caption{\label{fig:Dyn_T5mK_B1T_vargap0_gap0_1p427R} Low-temperature dynamics without collisions for the dimer with anisotropic response ($B = 1$ T). (a-c) Study varying the initial gap. (a) reduced-mass coordinate. (b) Number of collisions. (c) Maximum contact force. (d-f) Idem Fig.~\ref{fig:Dyn_T5mK_B1T_e1_wl49p81um} when the dimer is initially tuned to $\lambda = 49.85 \mu$m and $g_0 = 1.427R$. (d) Net forces on the system. (e) Velocities. (f) Particle trajectories.}
\end{figure}
Fig.~\ref{fig:Dyn_T5mK_B1T_vargap0_gap0_1p427R}  summarizes several runs that were performed with the aim to predicting dimers by avoiding collision events, as far as the collisions are understood as a cause of dimer breaking. Panel (a) shows the average absolute value of the reduced-mass coordinate as a function of the initial gap. This curve shows alternative peaks and valleys, which mean that several minimal ("orbits around $z$" or) inter-particle distances are obtained when collisions are less or not produced at all, which keeps the dimer assembled for low temperatures. (b) shows the collision events and (c) their resultant maxima in contact forces. The normal restitution coefficient represents elastic collisions, i.e., $e = 1$. As panels (b-c) are in clear correspondence, they show that the collisions cause the dimer breaking. As a result of the helical movement, the $x$ and $y$ binding components are 90-deg. out of phase in the first moments, and there is a net radiation pressure for the system along $z$, panel (d), as previously predicted in \cite{abraham-ekeroth_numerical_2023} for MO dimers of n-InSb. This is the only force component which remains as time runs, because the low temperature of this example does not cause appreciable thermal movements. As a consequence, all the components of the velocities level out with the exception of the $z$ components, (e), which increase linearly with time. Panel (f) illustrates the trajectories for the first time values. Noteworthy, the dynamics presented is exclusively obtained with the magneto-optical coupling between the particles. Fig S1 in the Supplementary Material (SM) gives another example for a different configuration and compares it with the case of single-like or non-interacting particles, to highlight the importance of the coupling phenomenon under the effect of collisions. 

\subsection*{Full dynamics at ambient temperatures}

\begin{figure}[h]
	\begin{centering}
		\includegraphics[width=9cm,keepaspectratio]{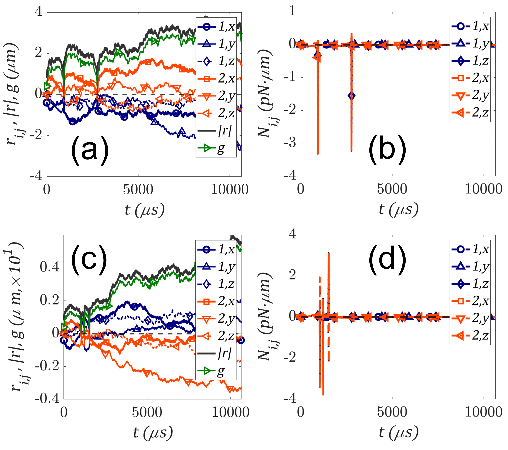}
		\par\end{centering}
	\caption{\label{fig:Dyn_20degC_B1T_gap0_2R_vs_e} Examples of dynamics for the dimer with isotropic n-InSb ($B$ off) with $g_0 = D$ at $T_0 = 20$ degC, $\lambda = 48.731 \mu$m, for two values of $e$. (a-b) [c-d] for $e=1$ [$e=0.5$]. (a,c) coordinate positions. (b,d) induced spin torques.}
\end{figure}

To explore the dynamics under brownian motion, the dimer with isotropic n-InSb ($B=0$) is firstly addressed by considering $e$ as a tuning parameter to stabilize the dimer, since attractive but unstable states were predicted from static studies for perpendicular configuration \cite{abraham-ekeroth_numerical_2023}. Here, two illustrative examples are shown in Fig.~\ref{fig:Dyn_20degC_B1T_gap0_2R_vs_e} with $e = 1$ (a,b) and $0.5$ (c,d) respectively. The positions and inter-particle distances can be seen in (a,c), and torques in (b,d) show the rotational inductions of the MO dimer with collisions. Though the parameter $e$ plays an important part in the dynamics through collision events, it seems that a stable dimer made of isotropic n-InSb cannot be formed at perpendicular alignment, even illuminating the system at the appropriate resonance wavelength. Quite clearly, all the magnitudes possess a strong variation when the collisions are detected. Concerning the rotational degrees of freedom, panels (b)-(d) show that the spin torques are strongly effected by the change that $e$ produces. The balance between collisions and brownian motion cannot stabilize the dimer at ambient temperature, no matter what value for $e$ is chosen. 

\begin{figure}[h]
	\begin{centering}
		\includegraphics[width=9cm,keepaspectratio]{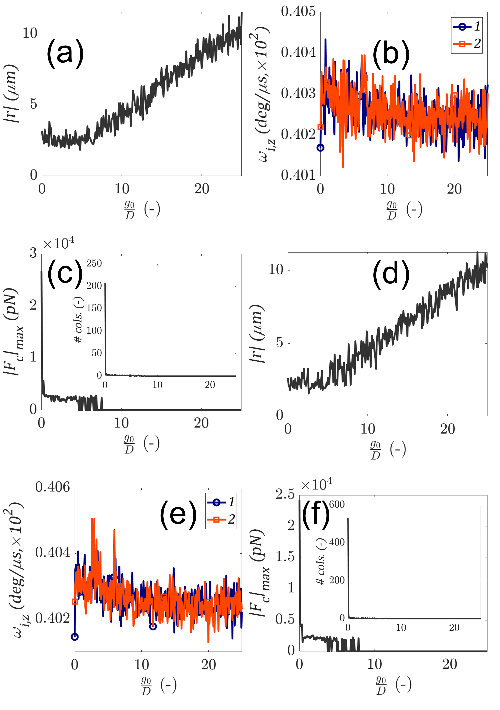}
		\par\end{centering}
	\caption{\label{fig:AvTimeDistribsGap0Var} Simulations varying the restitution coefficient $e$, at $T_0 = 20$ degC, $B = 1$ T, $\lambda = 49.85 \mu$m, as a function of the initial gap (average magnitudes). (a-c) [d-f] for  $e=1$ [$e=0.5$]. (a,d) reduced-mass coordinate. (b,e) angular velocities. (c,f) Contact forces and number of collisions (inset).}
\end{figure}

It is maybe worth mentioning that $e=0.5$ was taken as an "extreme" value to make deductions on a wide variation for $e$, but in principle such values could serve to model realistic situations where the particles are functionalized. Other conditions for the surrounding media could affect the performance of the particles concerning their collisions, e.g. their PH, colloidal dispersions (zeta-potential effects), viscosity, temperature, etc. 
Having recognized $e$ as a fundamental parameter that can modify the dynamics, the study now focuses on an dimer with anisotropic n-InSb ($B$ on) as a function of the initial gap, Fig.~\ref{fig:AvTimeDistribsGap0Var}. In particular, this figure compares two sets of simulations; (a-c) for $e=1$, and (d-f) for $e=0.5$. Based on the previous results, the illumination wavelength is $\lambda = 49.85 \mu$m, which can excite SPhP resonances in the dimers for a wide range of values $g_0$ under $B=1$ T.

Logically, the minimal values for $|r|$ occur at small initial gaps; however, Fig.~\ref{fig:AvTimeDistribsGap0Var}a shows an absolute minimum that corresponds to a value around 2-4 times $g_0/D$ and repeats along many simulation runs. This can be distinguished even when the fluctuations produced by thermal noise are moderate. The collisions events are repetitive at very small initial gaps (c and inset) so that one could think the dimer's stability is compromised in this region. On the other hand, the spin angular velocities are relatively high with a moderate level of fluctuations but neither are so sensitive to the gap variations nor the value for $e$. Actually, $e$ cannot influence the spins under the assumed model; it can be shown by using Eq.~\ref{eq:Jn} in Eq.~\ref{eq:SlidingFriction} that $e$ cancels out from the sliding friction coefficient. In agreement with Fig.~\ref{fig:AvTimeDistribsGap0Var}a, panel (c) presents strong contact forces with a moderate number of collisions for small initial gaps (inset). The resultant collision number is also related to the fact that these collisions are elastic. 

In the case of inelastic collisions, Figs.~\ref{fig:AvTimeDistribsGap0Var}d-f, it seems that the minimum binding distance, $|r|$, is given for zero initial gap or close to it (d). As some of the mechanical energy is lost through the collisions (besides the always existing optical absorption), these events are no so effective in breaking the dimer; the angular velocities do not seem to change on average (e) and the amount of collisions is higher with slightly weaker contact forces (f and inset), which means the dimer is more likely to remain stable when the gaps are small. As a whole, panels (a) and (d) show the same trend, meaning that the dimer is stable for small initial gaps no matter the value for $e$ is used. Though, $e$ influences some details of the movement such as the contact forces, and the amount of collisions. Such dependence is important in the design of mesoscale motors and nano or molecular factories \cite{valadares_catalytic_2010,raziman_optical_2015}.

\begin{figure}[h]
	\begin{centering}
		\includegraphics[width=9cm,keepaspectratio]{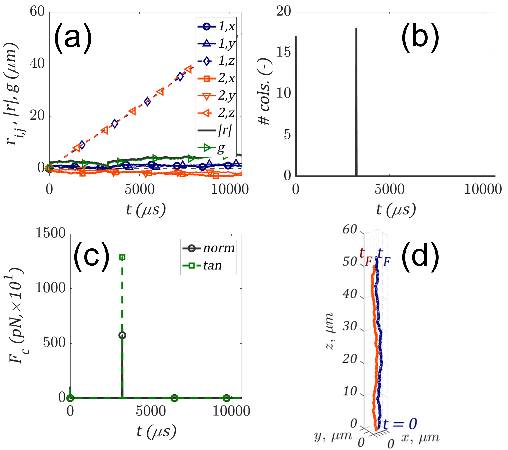}
		\par\end{centering}
	\caption{\label{fig:Dyn_T20degC_B1T_gap0_0p1R_e1_49p85um} Example of dynamics at $T_0 = 20$ degC when $B = 1$ T, $\lambda = 49.85 \mu$m, $e = 1$, and the initial gap fixed at $g = 0.1R$. (a) Relative positions. (b) Collision events. (c) Contact forces. (d) Perspective view showing the stable dimer's formation and its backward movement along $z$.}
\end{figure}

\begin{figure}[h]
	\begin{centering}
		\includegraphics[width=9cm,keepaspectratio]{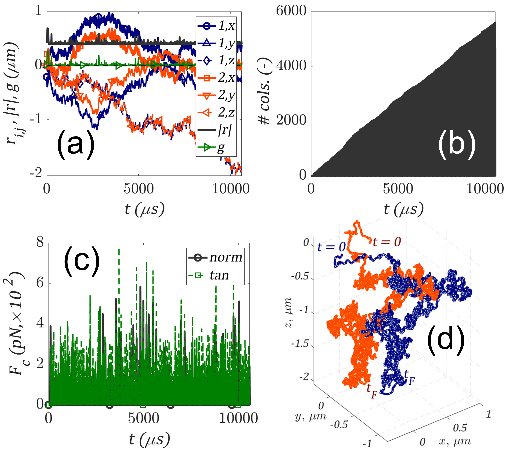}
		\par\end{centering}
	\caption{\label{fig:Dyn_20degC_B1T_gap0_0p1R_e1_72um} Idem Fig.~\ref{fig:Dyn_T20degC_B1T_gap0_0p1R_e1_49p85um} when $\lambda = 72.784 \mu$m to initially excite the lowest-energy SPP of the dimer. (a) Relative positions. (b) Collision events. (c) Contact forces. (d) Perspective view showing the stable dimer's formation and its backward movement along $z$.}
\end{figure}

With the purpose of illustrating the stability of the dimer when $B$ is on and as a complement to Fig.~\ref{fig:AvTimeDistribsGap0Var}(a-c), some results for single time-domain simulations are shown in Fig.~\ref{fig:Dyn_T20degC_B1T_gap0_0p1R_e1_49p85um}. This example deals with a dimer excited at its SPhP resonance, suffering elastic collisions and initially with a small gap, $g_0 = 0.1R$. As clearly seen from Fig.~\ref{fig:Dyn_T20degC_B1T_gap0_0p1R_e1_49p85um}a, the gap (or the center-to-center distance $|r|$) remains steadily with a net movement along $z$ due to radiation pressure. There are just a few collisions occurring at very different times (b), which means the particles stay close and forming a stable dimer after a long time. Here, the contact forces are specified by components (c), which help to understand the phenomena of collisions. The maxima are reduced compared to the previous figures; and in the second set of collisions (around $3233\mu$s) the tangential forces result much higher than the normal force components. Panel (d) shows a perspective view of the trajectory for the first instants of the movement. To explore more on the behavior of other variables and their inherent fluctuations, Fig.~S2 was added to the SM document.
The "MO binding + collisions" effect is not constrained to the SPhP and will behave differently for other resonant configurations, because the involved phenomena is very complex. To convey this idea, Fig.~\ref{fig:Dyn_20degC_B1T_gap0_0p1R_e1_72um} refers a time-domain simulation with elastic collisions and the same initial configuration $g_0 = 0.1R$ as in Fig.~\ref{fig:Dyn_T20degC_B1T_gap0_0p1R_e1_49p85um} but this time the lowest-energy SPP is excited, namely, $\lambda = 72.784 \mu$m. If one calculates "statically" the binding force for this example at the gap $g_0$, it gives repulsion. However, it was reported (for gaps $g=2R$) that the binding force $\Delta$ changes from negative to positive when the azimuthal angle (between the dimer's axis and $x$) varies periodically 90 deg., see Ref.~\cite{abraham-ekeroth_numerical_2023} for more details. Hence, a slight variation due to any type of force (optical, thermal, drag, etc.) may again lead to attraction and eventually colliding particles. In fact, this is what the example in Fig.~\ref{fig:Dyn_20degC_B1T_gap0_0p1R_e1_72um} shows; as a result of all the complex phenomena, the particles keep each other very close (a), the number of collisions is huge and continuously occurring at all times (b), and so both tangential and normal contact forces. Then, it turns out that the dimer is stable though it moves with modest variations because of the thermal noise (d). Moreover, this example shows negative radiation pressure, as can be seen from (d) when tracking the trajectories from $t_0=0$ to $t=t_F$. There is a small value for the net radiation pressure in the negative $z$ direction for the dimer that produces a notorious effect if the time simulated is long enough. For more studies about this example of excitation at the SPP, the reader is referred to the SM, where Fig. S3 summarizes time-average results with/out variation of the initial gap.

\begin{figure}[ht]
	\begin{centering}
		\includegraphics[width=9cm,keepaspectratio]{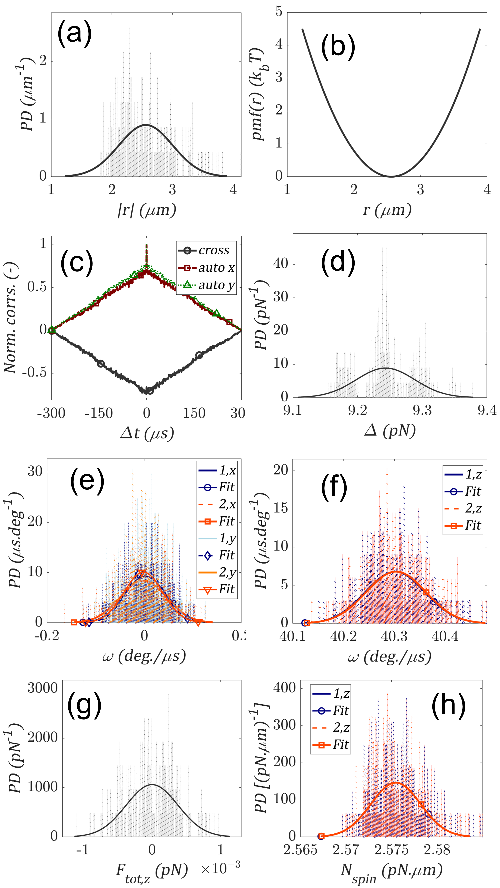}
		\par\end{centering}
	\caption{\label{fig:AvTimeDistribsGap0Fixed} Histograms and gaussian probability distributions for the essential time-average variables at $T_0=20$ degC when $B = 1$ T, $\lambda = 49.85 \mu$m (SPhP), and the initial gap is $g_0=1.59D=3.18R$, which lies around the minimum found in Fig.~\ref{fig:AvTimeDistribsGap0Var}a. (a) reduced-mass coordinate. (b) potential of mean force obtained from the fit in (a). (c) auto and cross correlations between the $x$ and $y$ components of the reduced-mass coordinate. (d) Binding force. (e,f) components of the angular velocities for each particle. (g) Radiation pressure for the system along $z$. (h) $z$-components of the spin torque for each particle.}
\end{figure}

Lastly, the distributions of the time-average magnitudes are assessed by assuming a constant value for the initial gap, Fig.~\ref{fig:AvTimeDistribsGap0Fixed}. The histograms represent normalized counts for a certain variable $\zeta$, and these former are fit by their respective, (normalized) gaussian probability densities, $P(\zeta)=\left(2\pi\sigma^2\right)^{-1/2}\cdot exp\left\{-0.5\left[\left(\zeta-\mu\right)/\sigma\right]^2\right\}$, with the appropriate average $\mu$, standard deviations $\sigma$, and physical dimensions. Such functions give a probability density for that variable assumed as continuous in the limit of infinite events. In the present, finite case, the histograms track $300$ runs for the dimer under a static field $B = 1$ T with simulation time up to $10$ ms. Elastic collisions are considered. Fig.~\ref{fig:AvTimeDistribsGap0Fixed}(a) shows the performance of the variable $r$ when its histogram is fitted with a gaussian probability $P(r)$. Such fit gives a mean value $\mu_r = 2.56034 \mu$m with a standard deviation $\sigma_r = 0.444706 \mu$m. The expectation value represents an average value for the gap of $\mu_g=\mu_r-D=2.1603 \mu$m which differs significantly from the initial gap, $g_0 = 1.59D = 0.6360 \mu$m in $1.5243 \mu$m. Such difference means that the MO particle-particle interaction with the effect of elastic collisions is essential in the dimer's formation. On the other hand, the effect of the thermal fluctuations can be estimated by $\sigma_r$. Panel (b) show the potential of mean force that is obtained from the fit $P(r)$ in Fig.~\ref{fig:AvTimeDistribsGap0Fixed}a as $pmf(r) =\, -k_BT \, ln\left[P\left(r\right)\right]$, which gives a kind of "effective" trapping potential (the arbitrary energy constant is set to zero so that $pmf(\mu_r)=0$) \cite{yan_three-dimensional_2012}. Moreover, assuming the region around the pmf's minimum as approximately harmonic and the validity of the equipartition theorem, namely, $\frac{1}{2}k_BT=\frac{1}{2}\kappa_\zeta \left\langle \zeta \right\rangle^2$, an estimation of the trapping stiffness $\kappa_\zeta$ for the each positional variable $\zeta=x,y,z,r,\int \vec{\omega}dt$ can be obtained. For brevity, this is not shown in the present manuscript.

Thus, the whole system responds to an average potential that defines the creation and control of the MO dimer (beware that the problem is non-conservative in general \cite{edelstein_magneto-optical_2021}). Other detailed information can be extracted from the correlations for the $x-y$ components of $r$ in Fig.~\ref{fig:AvTimeDistribsGap0Fixed}c. The analysis of correlations in the context of Langevin dynamics was reported as a robust method to indirectly measure optomechanical variables \cite{volpe_torque_2006,volpe_brownian_2007}, properties of out-of-equilibrium dynanics, to scan trapping potentials, calibrate photonic force microscopes \cite{bar-ziv_localized_1997}, or even design sensitive pressure-sensors in low-pressure setups \cite{arita_coherent_2020}. The auto-correlation measures the rate at which the dimer's motion loses coherence because of thermal fluctuations. In other words, the system loses "memory" to track the deterministic trajectory that would follow if no thermal noise was present. Then, the correlations also take the collision events into account. The shape of such correlations also gives information; although it was seen that the trajectories follow helical shapes, a clear restoring effective force prevails over the dominance of rotational effects, as the curves in (c) do not present sinusoidal or periodic oscillations beyond those generated by noise \cite{volpe_torque_2006,chand_emergence_2023}. Nevertheless, the negative cross correlation between $x$ and $y$ components indicates the presence of rotational motion for the dimer \cite{reichert_hydrodynamic_2004,herrera-velarde_hydrodynamic_2013}. On the whole, as the correlations decrease to zero in the order of 300 $\mu $s, the restoring forces are relatively strong compared to the noise forces and keep some memory of the deterministic movement. Accordingly, the binding-force fit reveals an average value for this example of $\mu_\Delta = 9.2432$ pN with a standard deviation of $\sigma_\Delta = 0.0451$ pN, Fig.~\ref{fig:AvTimeDistribsGap0Fixed}(d), the latter expressing a weak influence of thermal fluctuations on this force. By symmetry, the $x-y$ components of the (spin) angular velocities (e) fluctuate around zero with good accurancy, i.e., $\sigma_{\omega,x,y} \rightarrow 0$. For $\omega_z$, (f), an average value $\mu_\omega = 40.301$ deg$/\mu$s is obtained for both particles, as was reported for a $z$-aligned magnetic field \cite{abraham-ekeroth_numerical_2023}. The standard deviation is $\sigma_\omega= 0.0590$ deg$/\mu$s for this example.

On the other hand, the average radiation pressure results practically zero (g); and the particle spin torques, (h), equal an average of $\mu_N = 2.5755$ pN$.\mu$m with high precision ($\sigma_N = 2.724 \times 10^{-3}$ pN$.\mu$m).

Note that in general, the histograms are well represented by their gaussian distributions. However, this is an only valid scenario when the dynamics is not dominated by collisions. When the collisions become drastic, the gaussian fit seems no longer suitable for the dynamic observables, and more complicated histograms may appear. Fig. S3(d-f) in the SM provides an extension of  Fig.~\ref{fig:Dyn_20degC_B1T_gap0_0p1R_e1_72um}'s study for $\lambda = 72.784 \mu$m, where clear deviations from the normal distributions can be seen concerning the MO binding phenomena.

The methodology developed is both versatile and agile, facilitating the rapid assessment of the realistic dynamics of small-scale systems from a classical standpoint. Generally, the formation of an MO dimer in the presence of static magnetic fields appears to need careful control of particle dispersion to maintain minimal initial gaps. 

\section*{Conclusion}

The present numerical framework addresses the Langevin dynamics problem for magneto-optical dimers, taking particle collisions into account. This efficient tool models the particles as dipoles with minimal loss of generality, provided that materials such as n-doped InSb involve long excitation wavelengths compared to the particle size and display sizable MO effects. The method’s capabilities are demonstrated by illuminating the particles with two counter-propagating plane waves carrying identical circular polarization, hence eliminating the need to account for gradient forces or light spots. 

Although assembling dimers made of isotropic materials without light spots seems challenging, a comprehensive study of dimer formation is beyond the scope of this paper.

The rotational dynamics is coupled to the MO binding between particles, resulting in large spin components. Nevertheless, the minor values found for the \textit{orbital} angular velocities, which are not displayed for the sake of brevity, corroborate a prior publication by one of the authors, implying a feeble spin-orbit interaction in this system.

Through illustrative examples at very low and ambient temperatures, the detailed dynamics of dimers, including kinematic variables and contact forces, shed light on interacting mesoscale clusters. This is particularly relevant for the synthesis of molecular assemblies for specific applications like building nano/micro robots, particle trapping, and sorting.

Collectively, the process of magneto-optical binding can be defined by several external parameters such as the illumination source, magnetic field, the embedding medium's constitution and temperature, initial flow speeds, and particle dilution, among others. Yet, it is equally responsive to internal modifications, such as intelligent particle functionalization, which can alter the collision properties. To sum up, the following goals were achieved: \\

\noindent{- Magneto-optical matter was proposed by studying the conditions to assemble dimers.} \\
\noindent{- A general model to assess MO matter was built, which is valid for arbitrary incident-wave energy, values of static magnetic field, type of particle-particle collisions, force types, initial kinetic conditions, and environmental thermodynamics, where such environment can be a flow or a static fluid.}\\
\noindent{- The effects of collisions in magneto-optical dimers were analyzed, given their significance in realistic particle-particle interactions, especially under the presence of a magnetic field, which can produce both attractive and centrifugal forces.} \\
\noindent{- For conciseness, the developed method was later constrained in the Results to show a few examples. For instance, neither the particles’ initial velocities nor air flows were considered in the present work, though the model can perfectly solve these problems.} \\

In particular, concerning the construction of magneto-optical matter, the extension to $N$ particles is both natural and immediate when employing the discrete dipole approximation method for optical forces. This novel form of matter can manifest even when moderate thermal fluctuations or internal degrees of freedom (such as particle spins) exist. This makes the phenomenon very complex yet robust. The community is encouraged to start MO matter studies in multi-particle scenarios.

\clearpage
\section*{Acknowledgments}

R.M.A-E. would like to thank D.T. for his continuous support and encouragement to complete this research. Also GROC, UJI, UNCPBA, CICPBA, and CONICET are sincerely acknowledged for providing the necessary office and time resources that significantly contributed to the completion of this research. R.M.A-E. and D.T. acknowledge the DYNAMO project (101046489), funded by the European Union. 

\section*{Additional Information}

\subsection*{Data Availability}

Data sets generated during the current study are available from the corresponding author on reasonable request.

\subsection*{Competing interests}

The authors declare no competing interests.

\end{document}